# How Surface Make-Up and Receding Electrokinetics Determine the Sign and Magnitude of Electrification at Water–Hydrophobe Interfaces?

Yinfeng Xu & Himanshu Mishra*

Environmental Science and Engineering (EnSE) Program, Biological and Environmental Science and Engineering (BESE) Division, King Abdullah University of Science and Technology (KAUST), Thuwal, 23955-6900, Kingdom of Saudi Arabia

Sustainable Food Security Center of Excellence, King Abdullah University of Science and Technology (KAUST), Thuwal 23955-6900, Saudi Arabia

Interfacial Lab (iLab), King Abdullah University of Science and Technology (KAUST), Thuwal 23955-6900, Saudi Arabia

*Correspondence: himanshu.mishra@kaust.edu.sa

**Word limit: 250**

**Abstract:** It has been widely reported that as water contacts hydrophobic materials such as air or hydrocarbons (liquid or solid), the interfaces acquire a negative charge. It is not entirely clear whether this occurs due to the nature of water, or the hydrophobe, or purely the interface. Here, we probe the effects of surface chemistry and the speed of liquid–solid contact formation and separation on electrification. Glass capillaries grafted with mixed self-assembled monolayers of octadecyltrichlorosilane (ODTS) and (3-aminopropy)triethoxysilane (APTES) were exploited. Water was drawn inside these capillaries from an electroneutral reservoir, and the excess charge carried by the pendant droplets, if any, was quantified using an electrometer with a 100 fC resolution. Depending on the APTES content, the surface charge density at the water–hydrophobe interface ranged from negative (for ODTS) to near-neutral (for APTES 2-sec-exposure followed by ODTS) to positive APTES(5s)–ODTS. Next, we probed the charge ($Q$) contributions of the following steps on the electrification: (i) Contact ($Q_1^n$): as a dry capillary enters the water reservoir; (ii) Liquid uptake ($Q_2^n$): as water is uptaken; (iii) Capillary lift ($Q_3^n$): the filled capillary is removed from the reservoir; and (iv) Liquid release ($Q_4^n$): releasing the liquid back into the reservoir. This revealed that the electrification during water uptake ($Q_2^n$) varied with the rate of release during the previous cycle ($Q_4^{n-1}$), and it did not depend on the uptake rate. We explain these findings based on the electrical double layer theory, electrokinetics, and charge conservation, advancing the current understanding of electrification.

**Introduction**

Water's interfaces with hydrophobic materials are ubiquitous and implicated in numerous natural processes and engineering solutions, such as cloud chemistry[1, 2], thundercloud charging[3], aerosols[4], water-repellent insects[5], mulching[6] and packaging[7] in agriculture, separation and purification[8-10], pipetting[11], micro and nanoscale fluidics[12, 13], and designer surfaces for drag reduction[14] and preventing cavitation damage[15], and repelling low surface tension liquids[16]. Therefore, a clear understanding of physical and chemical phenomena at such interfaces is desirable. It has been widely reported that when water contacts a hydrophobic phase (solid, liquid, or gas), defined based on apparent contact angles, $\theta_o \geq 90°$,[17] the interface acquires a negative charge. This induces a positive charge onto the bulk water adjacent to the interface – consequently, pendant droplets formed at the end of a hydrophobic capillary carry a net positive charge[18, 19]. This observation has also been connected to the electrophoretic mobility of bubbles and oil droplets, leading to the generalization for electrification of water–hydrophobe interfaces[20-24]. A lot of experimental, theoretical, and computational effort has been invested in unraveling the nature of this negative charge during the last two decades. A number of mechanisms – often contradictory and intensely debated – have been reported the interfacial adsorption of water's intrinsic ions (i.e., hydroxide or protons)[25-32], deprotonation of surface groups[33, 34], faster autolysis of water at hydrophobic interfaces[22], partial charge transfer between water molecules[35] and water molecules and hydrophobes[36], dipolar organization of water[37, 38], airborne surfactants (i.e., impurities)[39-41], surface adsorption of bicarbonate ions due to ambient $CO_2$[42], volta potential[43], and cryptoelectrons[44]. In addition to surface-specific spectroscopic measurements, which typically utilize concentrated salty or acid/base solutions,[45] direct surface force measurements[33, 46, 47] and droplet charge measurements via electrometers and droplet deflection inside uniform electric fields have been reported[18, 43, 48, 49]. Complementary experiments with impinging or sliding droplets onto hydrophobic dielectrics with underlying metal electrodes have laid the foundation of droplet-based electricity generators and shed light on the effects of surface chemistry, stiffness, and discontinuous flows[18, 50-61]. Most of these experiments are, however, confined to a relatively narrow velocity range. Sliding droplets generally exhibit velocities in the range of 0.1–0.5 m/s, while spreading and retraction of the droplet upon impact yield contact line velocities of 0.3–3 m/s. A careful assessment of the effects of speeds of formation of liquid–solid contact, followed by separation, on electrification is warranted.

Some three decades ago, Faubel & Steiner noted that the speed of liquids moving on a solid surface critically influences electrification by shooting high-speed water jets in a low pressure chamber[62]. Building on this observation, Saykally & co-workers showed $H_2$ production from electrokinetic flows (at speeds of 50–400 m·s$^{-1}$)[63] and provided a mechanistic framework based on a modified Poisson–Boltzmann theory and continuum hydrodynamics, demonstrating that streaming currents similarly increase with velocity due to the coupling between fluid flow and the ion distribution in the electric double layer[64, 65]. Most recently, Ratschow et al. (2024) proposed a theoretical model for dewetting-induced charge separation in sliding droplets, and their results predict a decrease in charge separation with increasing velocity due to expansion of the diffuse double layer at the contact line[66]. On the slower end of the spectrum, viz., everyday pipetting (0.1–30 cm-s$^{-1}$), our team has unveiled a new mechanism for water's electrification based on the foundational work of Bard & co-workers[44, 67] and Lowell[68, 69]. Turns out, when water is withdrawn and dispensed from an electroneutral reservoir via a common hydrophobic capillary, e.g., polypropylene pipette or a perfluorodecyl trichlorosilane-coated (FDTS-coated) glass capillary, the droplet acquires a net positive charge and the reservoir acquires a net negative charge[70]. That is, common hydrophobic materials are negatively charged even in the air – a hitherto unappreciated fact – and the abovementioned mechanisms may have a secondary contribution, if any.

In this report, we address the following fundamental questions using a capillary and a pendant drop formed at its inlet:

(i) Could the sign and magnitude of electrification at water–hydrophobe interfaces be systematically modulated – from negative to neutral to positive – via surface chemistry (while exhibiting hydrophobicity)?
(ii) How does the electrification depend on liquid velocity, i.e., speed and direction of liquid uptake/release, as it enters/leaves the capillary?
(iii) Can we reconcile electrification with microscopic ion interactions at the interface and overall charge conservation?

We reveal that it is possible to achieve water–hydrophobe interfaces that are charge neutral or positively charged, i.e., charge reversal, via surface chemistry. Next, by varying liquid flow velocities over three orders of magnitude, we point out the key drivers for electrification during pipetting experiments.

**Results**

**Surface Make-up and Charge Reversal.** To study water's electrification on hydrophobic surfaces, we chose borosilicate glass capillaries for the ease of modifying their surface chemistry. The surfaces were first activated via a piranha treatment, followed by silanization with octadecyltrichlorosiliane (ODTS) in a toluene solution (protocols reported previously[9]). Atomic force microscopy and contact angle goniometry on flat $SiO_2$/Si wafers, as surrogates for glass capillaries, revealed a surface roughness < 2nm and water's advancing/receding contact angles of $\theta_A \geq 110°$ and $\theta_R \geq 90°$, respectively (SI Note S1). The water–ODTS interface, like other water–hydrocarbon interfaces, flaunts a negative surface charge density[58, 70]. To neutralize and even reverse this charge, we followed the beautiful work of Butt & co-workers[71].

Utilizing (3-aminopropyl)triethoxysilane (APTES), with p$K_b$ = 9.5,[72] activated glass surfaces were silanized via liquid-phase silanization for a tightly controlled duration, which was then followed by silanization with ODTS. Thus, glass surfaces with the following three silanization treatments were produced: (i) **ODTS**, (ii) **APTES (2 sec) followed by ODTS** (hereafter referred to as APTES(2s)–ODTS), and (iii) **APTES(5s)–ODTS** (Methods). Please refer to the advancing/receding contact angle measurements on these surfaces, attesting[17] to their hydrophobicity, in SI Note 1.

Next, we probed their electrification with water. Using the glass capillaries of the three treatments described above, water (50 μL, MilliQ Advantage 10 set up | resistivity 18.2 MΩ cm) was withdrawn from an electroneutral reservoir and dispensed into a Faraday cup connected to an ultrasensitive electrometer with a limit of detection of 0.1 pC (Figure 1a). The uptake and release of liquids were precisely controlled using a syringe pump operating at 10 mL-min$^{-1}$, enabling consistent fluid handling throughout the process. As the water droplet landed in the Faraday cup, the electrometer recorded a sharp jump in the measured charge, and the signal stabilized within 2 seconds (Figure 1b). Charge measurements for all surface treatments were repeated $n \geqslant 10$ times, and the results are presented as the means with error bars ascribing to one standard deviation (Figure 1c). For the capillaries coated with ODTS, the measured charged of the dispensed water at 10 ml-min$^{-1}$ was +0.50 ± 0.03 nC. In contrast, for the mixed SAMs, i.e., APTES(2s)–ODTS and APTES(5s)–ODTS, the net charges carried by the water droplets were +0.18 ± 0.02 nC and -0.42 ± 0.02 nC, respectively. This apparent charge reversal of water–hydrophobe interfaces demonstrates that electrification is largely governed by the hydrophobe, i.e., not by water, and it can be tuned via surface engineering.

By assuming that the surface charge for the ODTS coating equals to the negative value of the charge carried by the water droplet, the surface charge density was estimated as $\sigma = -q/A$, where $q$ is the charge of dispensed water and $A$ is the water-solid interfacial area. This is how we quantified the surface charge density of ODTS, $\sigma$ =-3.1×10$^{-6}$ C-m$^{-2}$, which was in reasonable agreement with the previously reported values of $\sigma$ = -4.5×10$^{-6}$ C m$^{-2}$ and $\sigma$ =-4.6×10$^{-6}$ C-m$^{-2}$ for pipette tips, presumably made of polypropylene by Choi & co-workers[18] and for perfluorodecyltrichlorosilane-coated (FDTS-coated) glass by Nauruzbayeva & co-workers[70] through similar experiments. We comment on the nature of charges of the capillary surfaces in the Discussion section.

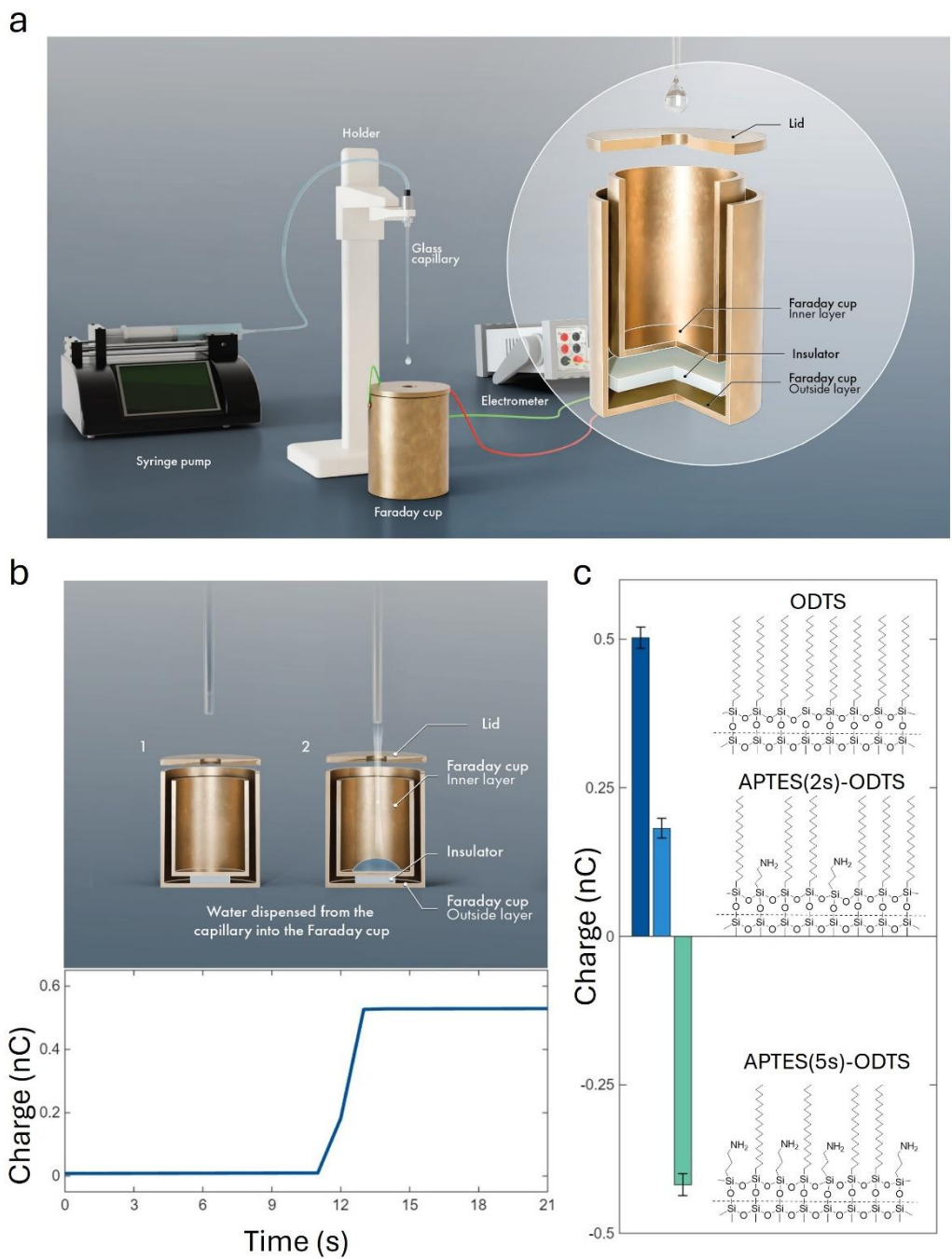

**Figure 1.** Direct measurement of 50 μL water droplets dispensed from hydrophobic capillaries at 10 ml-min$^{-1}$. (a) Experimental set up comprised a copper Faraday cup connected to an ultrasensitive electrometer (limit of detection: 0.1 pC). (b) In a typical experiment, water is withdrawn from an electroneutral reservoir, using a capillary of controlled surface treatment, and then dispensed as a single drop into the Faraday cup. An example is shown – as a droplet lands into the Faraday cup, charge is recorded. (c) Representative measurements for the following chemical make-ups: (i) ODTS, (ii) APTES(2s)–ODTS, and (iii) APTES(5s)–ODTS. Charge reversal observed with the increasing APTES concentration. Note: error bars represent one standard deviation.

**Electrification During Liquid Uptake and Release Steps**

Next, we interrogated the effects of the rates of uptake and release of water (50 μL) on charge carried by the released droplets. As the volumetric flow rates were varied from 1–100 ml-min$^{-1}$, so did the speeds of the waterfront advancing/receding (i.e., wetting/dewetting) within the capillaries. To gain deeper insight, we decomposed each water uptake/release cycles from/into a water reservoir (placed inside the Faraday cup) into four stages (Figure 2a), and quantified the corresponding charges: **(i) Contact ($Q_1^n$)** as a dry capillary enters the water reservoir; **(ii) Liquid uptake ($Q_2^n$)** as water is uptaken into the capillary; **(iii) Capillary lift ($Q_3^n$)** the filled capillary is removed from the reservoir (and the electrometer); and **(iv) Liquid release ($Q_4^n$)** releasing the liquid back into the reservoir. Here, $Q_i^n$ denotes the net charge recorded by the electrometer during stage $i$, and the superscript *n* indicates the *n*-th test in a series of consecutive tests. A positive sign indicates a net gain of positive charge at the electrometer. All charge data were obtained following application of drift correction. For a more detailed description of the four stages and the procedure used for measurement drift correction, see SI Note 2.

The experimental results revealed that for all the surface treatments and all the uptake/release rates, the sum of the charges measured across a cycle – water uptake and release – were centered at zero (Figure 2b), i.e., total charge was conserved. Because only a small portion of the capillary – the tip – contacted the water reservoir during the contact/lift stages, the charges measured in $Q_1^n$ and $Q_3^n$ were significantly smaller than $Q_2^n$ and $Q_4^n$, typically within ±1 nC. These observations establish that the charge transferred during the Liquid Uptake step was equal in magnitude and opposite in sign to that during the Liquid Release step, i.e., consistent with charge balance/conservation.

Crucially, we discovered that the release rate of water had the most significant influence on the charge carried by the released droplet ($Q_4^n$). For instance, as the dispensing flow rate was varied from 1 to 100 ml-min$^{-1}$, the measured charge from the ODTS-coated capillaries increased from 0.43 ± 0.03 nC to 0.67 ± 0.05 nC (Figure 2c); for the APTES(2 s)–ODTS capillaries, the charge increased from 0.15 ± 0.01 nC to 0.27 ± 0.02 nC (Figure 2d); and for the APTES(5s)–ODTS capillaries the charges increased from -0.31 ± 0.02 nC to -0.55 ± 0.02 nC (Figure 2e). In essence, as the speed of receding water meniscus increased, so did the water–hydrophobe electrification. Against our expectation, we discovered that the rate of liquid uptake, which is merely the opposite of the release rate, did not have significant effect on the charge accumulation $Q_4^n$ (Figure 2f–h). Lastly, the effects of the contact and capillary lift stages were found to be minimal on $Q_2^n$ or $Q_4^n$.

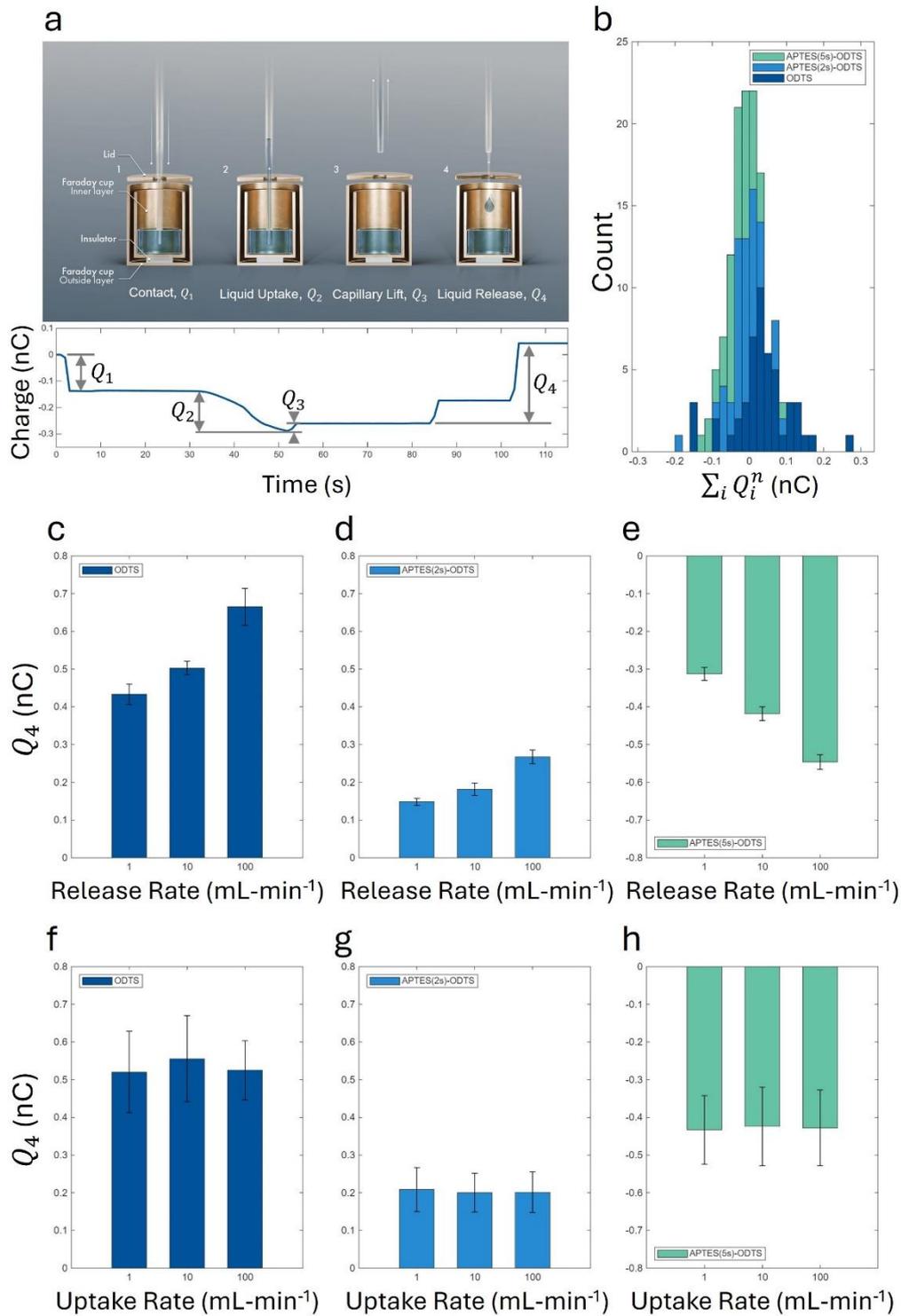

**Figure 2.** Charge measurements as a function of uptake and release rates of 50 µL water in and from the hydrophobic capillaries described in Figure 1. **(a)** Illustrates four major stages for liquid dispensing: contact, liquid uptake, capillary lift and liquid release, with the corresponding charges. The charge curve shown corresponds to a representative ODTS-coated capillary cycle, with both uptake and release rates set to 10 mL-min$^{-1}$; the data have been corrected for drift. **(b)** Charge neutrality for each cycle. The net charge $\sum_{i=1}^{4} Q_i^n$ is statistically indistinguishable from zero across all flow rates and surface chemistries. **(c-e)** Charge $Q_4^n$ measured during Liquid Release as a function of the release rate for capillaries coated with **(c)** ODTS, **(d)** APTES(2s)–ODTS, and **(e)** APTES(5s)–ODTS, respectively, showing an irrefutable effect. **(f-h)** As the liquid uptake rate is increased, the measured $Q_4^n$ shows no significant difference, for the capillaries.

**Effects of Contact Duration on Electrification**

As the release rate critically influences electrification (Figure 2c–e), we inquired whether it is the lifetime of the liquid–solid contact or the speed of the liquid the key factor. To assess whether the contact duration contributes to electrification, a controlled experiment was conducted in which the delay between uptake and release was varied from 3 s to 30 s, keeping the uptake and release rates constant. The net charge measured in each case exhibited no statistically significant difference, indicating that the variation in charging is not attributable to the contact time (Figure 3a).

**Electrokinetics: Effects of the Speed of Outgoing Liquid Meniscus**

Next, we quantified the effects of liquid release rates on electrification, i.e., the speed and acceleration of the receding liquid meniscus. High-speed imaging was used to measure the velocity and acceleration through curve fitting (see Methods and SI Note S3). As the rates of uptake/release were altered from 1 to 10 to 100 ml-min$^{-1}$, the maximum speeds and accelerations of the waterfronts increased in the range of 22–360 mm-s$^{-1}$ and 80–3000 mm-s$^{-2}$, respectively. Notably, a key distinction between our experiments and prior literature on streaming currents[62, 63, 65] lies in the fact that the fluid volume is finite, rendering the flow non-steady, i.e., speed increases from zero to $V_{max}$ over time. Under the same release rate, the velocity and acceleration profiles are generally consistent across capillaries with different surface modifications (Figure S6). Therefore, we use the maximum velocity $V_{max}$ and maximum acceleration $A_{max}$ as representative parameters to characterize the flow motion, with their values under different release rates listed in Table S2. The measured charge exhibits a quasi-linear dependence with respect to $V_{max}$ and $A_{max}$ on the semi-log scale, suggesting a power-law relationship (Figures 3b–c).

To avoid misinterpretation, we note that the observed rate dependence is not classical streaming current, considering that during pipetting the liquid column inside the capillary is decreasing in volume, such that a steady state is never achieved. We elaborate on this point in the Discussion.

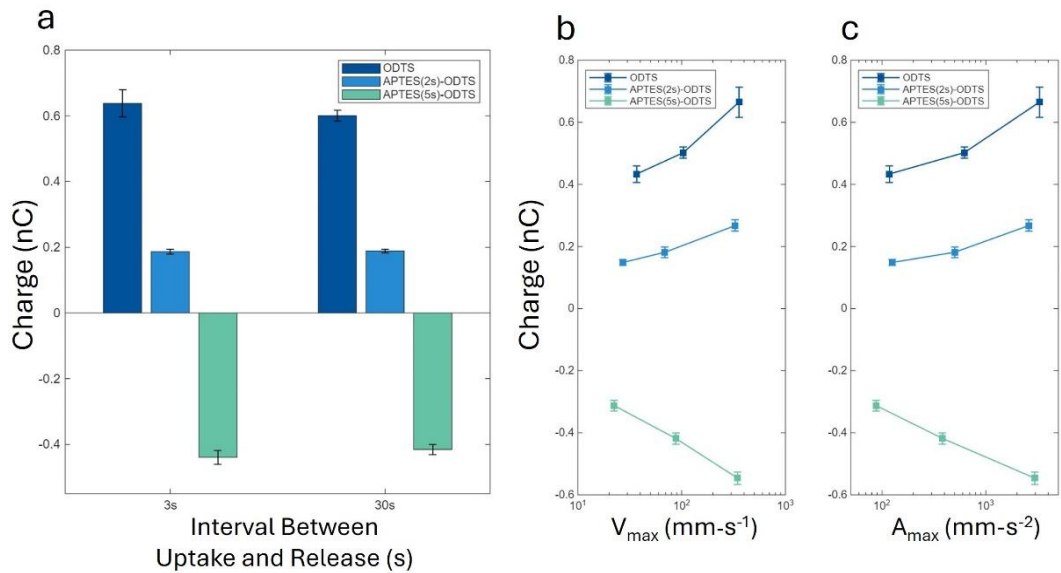

**Figure 3. (a)** Charge measured with different time intervals—3 s and 30 s—between liquid uptake and release. **(b)** Charge as a semi-logarithmic function of the maximum velocity during liquid release. **(c)** Charge as a function of the maximum acceleration during liquid release.

**Charge Conservation.** We conducted a series of continuous experiments involving three distinct uptake/release rate combinations, ranging from 10-100 ml-min$^{-1}$. For each combination, 3–5 successive trials were performed without interruption. As shown in Figure 4a, when both uptake and release rates were set to 10 μL s$^{−1}$ initially, the net charge remained stable across trials. Increasing the uptake rate while keeping the release rate constant had little effect on the measured charge in either stage. When the uptake rate was reset to 10 μL s$^{−1}$ and the release rate increased to 100 μL s$^{−1}$, the charge during first uptake remained largely unchanged, but the release stage showed a significant increase in net charge.

Repeating this cycle led to a reproducible pattern in which the charge during uptake became nearly equal in magnitude and opposite in sign to that during the preceding release. This behavior suggests a dynamic equilibration process, where charge transferred during uptake offsets that during release. Accordingly, the experimental results and charge conservation behavior presented in Figure 4a can be summarized by the conservation relation:

$$Q_2^n = -Q_4^{n-1}$$

To verify the robustness of the relationship between charging associated with liquid release ($Q_4^{n-1}$) and liquid uptake ($Q_2^n$) in sequential cycles, we retrospectively analyzed all over 113 datasets collected in our laboratory. The scatter plot reveals a Pearson's correlation coefficient of –0.99 with a highly significant p-value of 4.94 × 10⁻¹⁰⁰, strongly validating the consistency of this conservation law (Figure 4b).

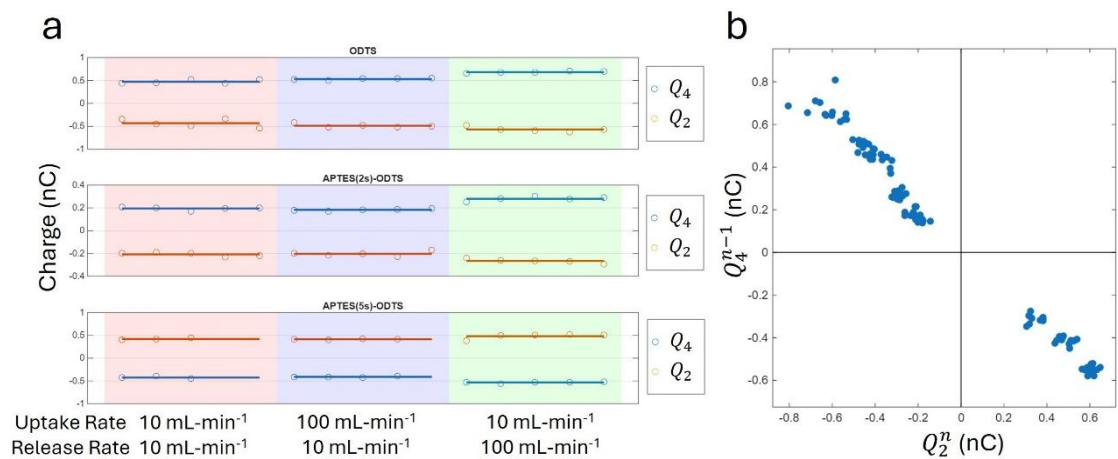

**Figure 4. (a)** Continuous experiments were conducted on three types of capillaries. Initially, several trials were carried out with both the uptake and release rates set to 10 mL-min⁻¹ (red background). Next, the uptake rate increased to 100 mL-min⁻¹ while the release rate remained unchanged (blue background). Finally, the uptake rate returned to 10 mL-min⁻¹ and the release rate was increased to 100 mL min⁻¹. Circles represent individual measurements, and the corresponding lines indicate the average values of the measurements. **(b)** Scatter plot of $Q_2^n$ versus $Q_4^{n-1}$ from 113 continuous measurements, demonstrating a strong negative correlation (Pearson coefficient = –0.99, p = 4.94 × 10⁻¹⁰⁰), consistent with the charge conservation relation.

**Discussion:**

Having established a clear flow-rate dependence of the electrification at water–hydrophobe interfaces, we now examine the underlying factors and mechanisms. In Figure 3b–c we observe that the variation of total charge scales approximately linearly with the logarithm of the maximum flow speed or acceleration, rather than linearly with speed as in classical streaming current. Prior studies on streaming currents have consistently demonstrated a near-linear positive correlation between fluid velocity (or applied pressure) and the resulting current[62, 65, 73, 74], a relationship well described by combining Poiseuille flow with the Poisson–Boltzmann electrical double layer theory. The key distinction in our experiment lies in the fact that the fluid volume is finite; therefore, only transient, non-steady flows exist. High speed imaging revealed that the instantaneous speeds and accelerations of the receding liquid-fronts inside the capillaries in real time ranged from 22–360 mm-s⁻¹ and 80–3,000 mm-s⁻², respectively (Movies S1–9, SI Note 3). Under these conditions, it can be proved that regardless of the electrical double layer and hydrodynamic flow profiles, assuming them to be decoupled, the total charge carried by the liquid remains constant (SI Note 4). In other words, although the instantaneous current may vary with the flow speed, the integrated total charge predicted by the classical streaming current framework remains constant – a result inconsistent with our experimental observations. Development of a quantitative model to explain these observations falls beyond the scope of this report.

As for charge conservation, based on the cross-cycle relation $Q_2^n = -Q_4^{n-1}$, we now examine the remaining contributions from Contact and Capillary Lift. Figure 5a summarizes the net charges measured at Contact ($Q_1^n$) and Capillary Lift ($Q_3^n$) for three surface treatments. Both $Q_1^n$ and $Q_3^n$ remain small in magnitude, substantially smaller than $Q_2^n$ and $Q_4^n$ (|0.1–0.6| nC). This result is consistent with the very limited liquid-solid contact area during the stages.

A notable feature is the deviation of $Q_1^n$ is large, and occasional sign inversions were observed. By contrast, $Q_3^n$ exhibits narrower spread and fewer sign changes. We interpret the unique feature of $Q_1^n$ as evidence that the interface immediately prior to contact retains a short-lived "memory" of the preceding release event, $Q_4^{n-1}$. During releases, the receding meniscus could transiently redistribute counter ions in the near-surface Stern layer, leaving a locally out-of-equilibrium interfacial environment. When contact is re-established with water, wetting of the capillary tip creates an electrical pathway to the inner wall of the Faraday cup. This brief grounding-like connection rapidly equalizes the residue charge, and as a result $Q_1^n$ is small on average but strongly case-dependent: its sign and magnitude are influenced by the interfacial charge left by the preceding release $Q_4^{n-1}$. The subsequent Liquid Uptake ($Q_2^n$) sustains and enlarges this electrical connection, promoting further equilibration of the interface and plausibly accounting for the comparatively stable $Q_3^n$ observed at Lift.

With cycles indexed by $n$, the conservation for all four stages writes:

$$Q_4^{n-1} + Q_1^n + Q_2^n + Q_3^n = 0,$$

which can be rearranged as:

$$Q_1^n + Q_3^n = -(Q_4^{n-1} + Q_2^n)$$

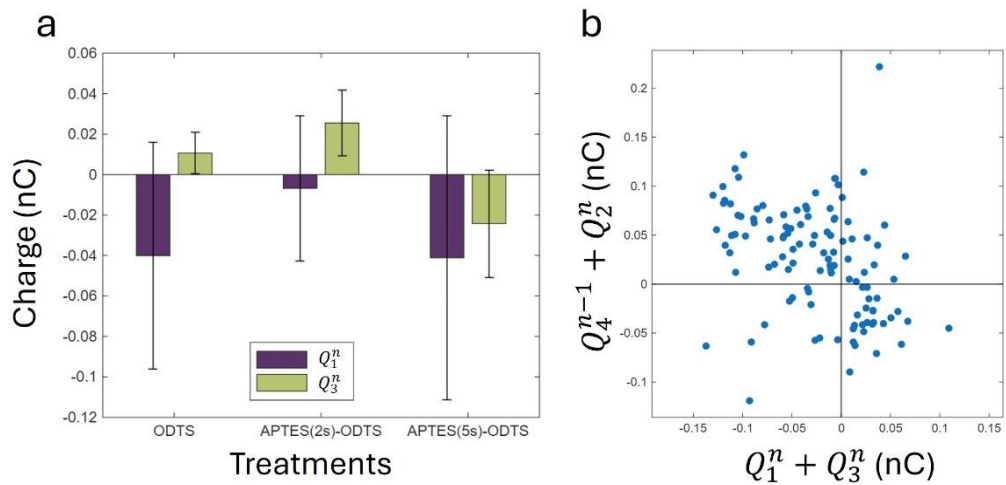

**Figure 5**. Validation of charge conservation relationships across consecutive tests. **(a)** Charges recorded during the Contact ($Q_1^n$) and Capillary Lift stages ($Q_3^n$). While both within ±0.1 nC, $Q_1^n$ shows a large standard deviation. **(b)** Scatter plot of $Q_1^n + Q_3^n$ versus $Q_4^{n-1} + Q_2^n$, showing a moderate but statistically significant negative correlation (Pearson coefficient = –0.38, p = 3.10 × 10$^{-5}$).

This reformulation explicitly reflects the quasi-symmetric nature of the two reciprocal process pairs—uptake versus release, and contact versus capillary lift—thereby enabling discussion of the overall charge conservation between these paired stages, with only minor residual imbalances. Notably, the absolute magnitudes of the summed terms are very small, typically within ±0.1 nC. Such uncertainties include a baseline variation of 2.3 ± 0.3 pC·s$^{-1}$ that may persist even after calibration (see SI Note 2 and Figure S2), underscoring the robustness of the observed conservation. As shown in Figure 5b, the scatter plot of ($Q_1^n + Q_3^n$) versus ($Q_4^{n-1} + Q_2^n$) yields a Pearson correlation coefficient of –0.38 (p = 3.10 × 10$^{-5}$). Although the negative correlation is moderate, its statistical significance supports this conservation-based interpretation. Importantly, the results further suggest that the charge at the contact stage ($Q_1^n$) is sensitive to the system's electrification history and, under substantial mismatch between uptake and release, may even undergo sign reversal. By contrast, the capillary lift stage ($Q_3^n$) is less affected, as the system becomes effectively "grounded" following liquid uptake.

Based on these insights, we provide a granular understanding of the various stages of electrification at water–hydrophobe interfaces in the context of pipetting. As established previously and reinformed here, common hydrophobic interfaces such as ODTS-coated surfaces naturally carry a certain amount of negative charge. This is presumably due to surface-bound electrons, and we refer the reader to prior work by Lowell & co-workers[68, 69, 75] and Bard & co-workers[44, 67]. As the negatively charged capillary begins to draw in water, excess H$_3$O$^+$ ions are drawn in forming an electric double layer. As the capillary is removed from the water reservoir, an equal and opposite negative charge is recorded. Now, when the water is dispensed gradually, i.e., at slow release rates, the tightly bound counterions on the surface may not be washed away by the liquid. Instead, they may remain electrostatically bound onto the solid surface with water – a patchy film or even a water cluster – adequate for ion hydration. A higher flow velocity, which corresponds to a greater energy input, may drive out the counterions due to stronger shearing forces, detaching them from the surface and resulting in a higher measured charge ($Q_4^{n-1}$). Due to this activity, the surface gets modified (or spring-loaded) to then attract more counterions when it encounters water the next time ($Q_2^n$). In the case of APTES(2s)–ODTS and APTES(5s)–ODTS, where amine groups were grafted prior to chemisorbing ODTS onto the surface, as water contacts, quaternary ammonium ions (–NH$_3^+$) form. Thus, the electrical double layer contains excess hydroxide ions, and during the water release process a net negative charge is measured. The magnitude of this charge is also influenced by the dispensing velocity along the abovementioned lines. These insights should be translatable to other Bronsted acid and bases as per their pH–p$K_a$ relationships. We hope that computational chemists will soon prove or disprove our viewpoint advancing the current understanding.

We close this section by commenting on the recently reported anomalous formation of $H_2O_2$ at water–hydrophobe interfaces, leading to an intense debate[25, 76-84]. For recent reviews on these interrelated topics, please see the references[32, 85-89]. Based on our work, we question the hypotheses based on the specific $OH^-$ ion adsorption at the water–hydrophobe interfaces. As revealed here, the nature of the solid surface and the flow conditions dominate the sign and magnitude of electrification. Decoupling the effects of instantaneous acceleration from instantaneous velocity on charging needs further investigation, and it may help design platforms for green chemistry[85], microscale energy harvesting[55, 61], and so on.

## Conclusion

This report demonstrates that water–hydrophobe interfacial charge (i.e., sign and magnitude) deduced by interrogating pendant droplets depends on the surface chemical make-up and receding electrokinetics. Experiments reveal that that common hydrocarbon surfaces such as ODTS are negatively charged, presumably due to surface-bound electrons, and this overall charge can be tuned to near-neutral or positive by grafting positively charged groups such as amines. Strong correlation between the magnitude of charge carried by pendant drops and the velocity and acceleration of the liquid–air interface during (prior) dispensing was observed. This indicates that as water recedes, it can alter the surface composition – the details of which necessitate further investigation. Quantitative insight into the multi-step pipetting process – contact, uptake, lift-off, and dispense – not only sheds light on the nature of charges implicated in water–hydrophobic electrification but will aid the rational design of droplet-based technologies.

**Materials and Methods**

**Sample Preparation.** Cylindrical capillaries (borosilicate glass, 1.5–1.8 mm × 100 mm, Kimble Chase, Cat. No. 34500-99) were rinsed with acetone and Milli-Q water, followed by a hydroxylation ("surface activation") with a fresh-made piranha solution ($H_2SO_4$: $H_2O_2$ = 4:1, v/v) at approximately 70°C for 30 minutes. After activation, samples were rinsed with Milli-Q water again and dried in a vacuum oven at room temperature for about 1 hour. Upon cooling, samples were subjected to one of three silanization treatments:

- Treatment 1 (ODTS only): Samples were immersed in trichloro(octadecyl)silane (ODTS) for 5 minutes.
- Treatment 2 (APTES(2s)- ODTS): Samples were briefly exposed to (3-aminopropy)-triethoxysilane (APTES) for 2 seconds, rinsed with toluene, and then immersed in ODTS solution for 5 minutes.
- Treatment 3 (APTES(5s)-ODTS): Similar to treatment 2, but with the APTES exposure time of 5 seconds.

All the silanizations were done in 20 mL fresh-made solution (toluene:silane = 400:1, v/v) at ~20°C. At the end they were rinsed by toluene and ethanol, dried in the vacuum oven and stored in sealed glass test tubes overnight before testing. Capillaries were glued to 20G*1" Monoject needles (Kendall) to allow connection to tubing or pipettes.

**Charge Measurement.** A Faraday cup made of brass were used for the measurement. It consists of an outer cylinder (8 cm diameter, 12 cm height) and an inner cylinder (6 cm diameter, 10 cm height), separated by dielectric foam. It was connected to an ultrasensitive electrometer (Keithley 6517B) with a detection limit of 10 fC (Figure 1). The capillaries were mounted on a holder which allows it to move up and down by hand and are connected to a PHD ULTRA™ syringe pump (Harvard Apparatus) with a PTFE capillary tube to enable precise liquid uptake and release.

Before measurements, the Faraday cup was connected to the electrometer and kept for at least overnight, to render the system stable. For each trial, 50 μL of Milli-Q water (≈pH 5.5) was taken into the capillary (total volume ~ 90 μL). The water was hold inside the glass capillaries only and never contacted with the tubing or syringe, avoiding unwanted contaminations. When the water was dispensed from the capillary into the Faraday cup, the transferred charge was captured and recorded using a LabVIEW software.

**Velocity Measurement.** A high-speed camera (Phantom® v1212, with a Nikon 55mm f/2.8 Micro Nikkor Lens) was used to film the dispense process with a resolution of 384×288 pixels. A backlit LED source enhanced the contrast. The frame rate was set between 100 to 3,000 fps depending on the dispensing rate. A ruler was used for calibration and the spatial resolution was determined to be 3.69 μm/pixel. Image sequences were analyzed using MATLAB. (SI Note 2)

**Conflicts of interest**

There are no conflicts to declare.

**Acknowledgements**

The authors gratefully acknowledge Mr. Moaz Kattoah and Prof. Sigurdur T. Thoroddsen from KAUST for providing an optical lens and an illumination system for high-speed imaging (Figure S3).

# How Surface Make-Up and Receding Electrokinetics Determine the Sign and Magnitude of Electrification at Water–Hydrophobe Interfaces?

Yinfeng Xu & Himanshu Mishra*

Environmental Science and Engineering (EnSE) Program, Biological and Environmental Science and Engineering (BESE) Division, King Abdullah University of Science and Technology (KAUST), Thuwal, 23955-6900, Kingdom of Saudi Arabia

Sustainable Food Security Center of Excellence, King Abdullah University of Science and Technology (KAUST), Thuwal 23955-6900, Saudi Arabia

Interfacial Lab (iLab), King Abdullah University of Science and Technology (KAUST), Thuwal 23955-6900, Saudi Arabia

*Correspondence: himanshu.mishra@kaust.edu.sa

---

**This PDF file includes:**

Supporting text:
Supplementary Note 1. Surface Characterization
Supplementary Note 2. Experiment Details for Uptake/Release Rate
Supplementary Note 3. High Speed Imaging and Post Processing
Supplementary Note 4. Limitations of Classical Streaming Current Models in Finite-Volume Capillary Flows

**Supplementary Note 1: Surface Characterization**

To characterize the surface quality, silicon wafers (Silicon Valley Microelectronics, p-type <100>) were cutting into pieces (1×6 cm$^2$) and underwent the same treatment as glass capillaries. On the wafer samples advancing/receding contact angles were measured (details below). Surface topography was characterized via AFM and RMS roughness was calculated (Table S1).

**Contact Angle Measurement.** Advancing and receding angles were tested with Krüss Drop Shape Analyzer DSA 100. The angles were measured by adding 10 µL to a 2 µL droplet and then removing it, all at a rate of 0.2 µL s$^{-1}$. For each wafer sample, at least three different points were tested and four measurements for advancing and receding angles were made during each advancing-receding cycle (Table S1). Recorded images were analyzed using *Advance* software (Krüss GmbH) with a manual baseline and 'tangent' fitting method for the droplet shape. The treated surfaces were found to be hydrophobic (Table S1).

**Surface Topography.** AFM imaging was performed using a JPK Nanowizard Ultraspeed-II in tapping mode with silicon probes (tip radius 8 nm, spring constant 2.8 N/m, frequency 75 kHz). Topography data were processed using Gwyddion software, including first order flattening and RMS roughness calculation (Figure S1).

**Table S1.** Contact angles and surface roughness for the samples.

| Coatings | Advancing Angle (°) | Receding Angle (°) | RMS Roughness (nm) |
|---|---|---|---|
| ODTS | 121 ± 1 | 92 ± 3 | 1.2 |
| APTES(2s)-ODTS | 117 ± 2 | 97 ± 2 | 1.7 |
| APTES(5s)-ODTS | 117 ± 2 | 96 ± 2 | 0.9 |

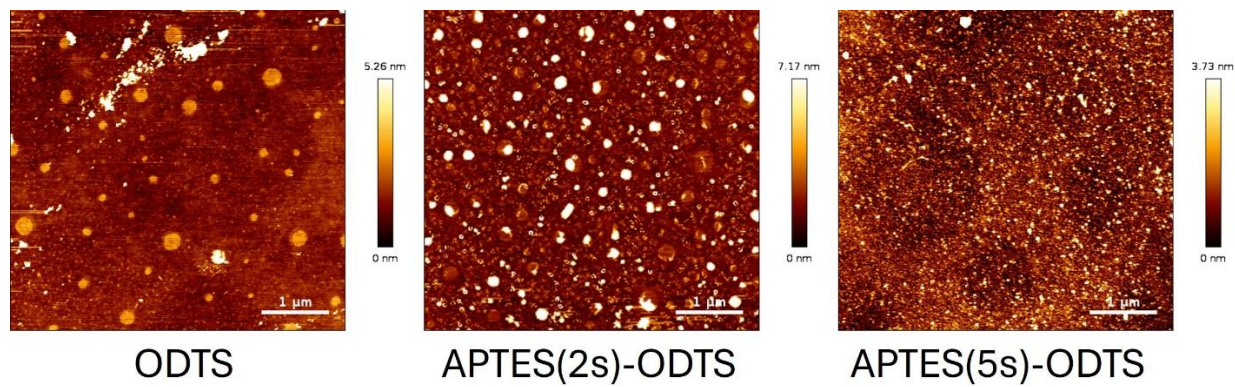

**Figure S1.** AFM imaging for wafer with three treatments.

**Supplementary Note 2: Experiment Details for Uptake/Release Rate**

To better understand the detailed charge transfer during the full liquid handling cycle, charge was monitored and analyzed across four key stages of the process (Figure S2a). In each experiment, water was initially placed in the Faraday cup. Figure S2b presents a representative charge transfer profile for an ODTS-coated capillary operated at a low flow rate of 1 mL min$^{-1}$ for both liquid uptake and release. The light blue curve represents the raw data, while the dark blue curve shows the data after drift correction.

Stage 1 - Contact, ($t_1$–$t_2$). The capillary was empty and brought into contact with the water surface in the Faraday cup. A small negative charge was detected, which is attributed to the naturally negative surface potential of the ODTS-treated capillary.

Stage 2 - Liquid Uptake, ($t_2$–$t_4$). 50 μL of water was drawn into the capillary by the syringe pump. During this period, the measured charge gradually became more negative, reflecting the accumulation of charge during capillary filling.

Stage 3 - Capillary Lift, ($t_4$–$t_5$). The capillary was manually lifted and suspended above the Faraday cup. A small positive charge was observed during this stage.

Stage 4 - Liquid Release, ($t_5$–$t_7$). Water was dispensed back into the Faraday cup at the same low rate of 1 mL min$^{-1}$. It took approximately 25 seconds to fully empty the capillary. Between $t_5$ and $t_6$, the charge signal remained stable with no major increase or decrease. This plateau is likely due to two factors: mechanical backlash in the syringe pump and the slow infusion rate. Water formed pendant droplets at the capillary tip, detaching once their weight exceeded the adhesion force. At the moment of droplet detachment (around $t_6$), a clear positive charge step was observed.

At this low release rate, the 50 μL volume typically formed two distinct droplets before the capillary was emptied, resulting in two charge steps visible between $t_6$ and $t_7$. At higher infusion rates, the entire volume was released at once, producing only one charge step during the release stage. The total duration of the liquid release stage varied significantly with infusion rate: approximately 40 seconds at 1 mL-min$^{-1}$, compared to about 20 seconds at 100 mL-min$^{-1}$ (including pump backlash delay).

**Drift and Correction.** A slow, consistent positive drift was observed in all raw charge data over measurements. This drift is likely due to electrochemical interactions between water and the brass walls of the Faraday cup, and the presence of dissolved oxygen and $CO_2$ under ambient conditions.

To correct for this drift, the period ($t_5$–$t_6$)—during which the capillary was held above the cup and no liquid transfer occurred—was used as a baseline reference. A linear fit of the charge signal over this interval was extrapolated across the entire dataset, and subtracted to yield the corrected charge curve (shown in dark blue in Figure S2b). This approach reliably removed background drift and was consistently applied across all measurements.

We analyzed the baseline drift across 149 experimental runs, including different capillary types, uptake rates and release rates. As shown in Figure S2c, the measured drift remains minimal throughout tests, with a mean value of 2.3 pC·s$^{-1}$ and a standard deviation of 0.3 pC·s$^{-1}$. This narrow spread indicates excellent consistency, suggesting that the drift is both small in magnitude and robust across experimental variations. These results validate the reliability of subsequent charge measurements and confirm that baseline drifts have a negligible impact on data interpretation.

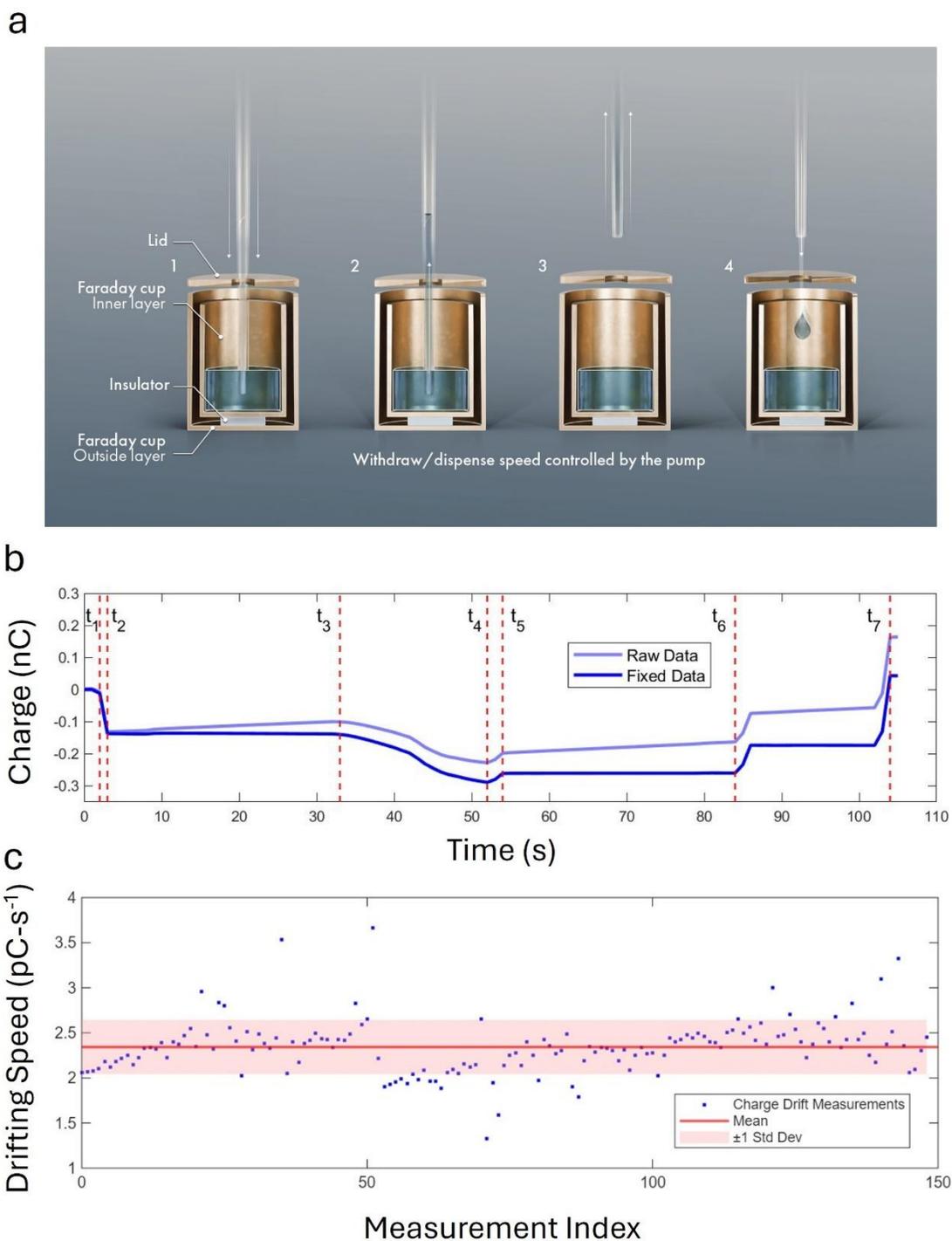

**Figure S2.** Charge measurement and drift correction. **(a)** Schematic of the four-stage liquid handling cycle (reproduced from Figure 2a in the main text). Shown here to facilitate stage-specific charge signal analysis presented in Figure S2b. **(b)** shows a typical result obtained during the four

steps. $t_1$-$t_2$: Contact, $t_2$-$t_4$: Liquid uptake, $t_4$-$t_5$: Capillary lift, and $t_5$-$t_7$: Liquid release. The light blue curve shows the raw data, which was corrected for drift to produce the dark blue 'fixed data' curve with accurate values for each stage. The sample shown in (b) is dispensed with an ODTS capillary at a low rate of 1 mL min$^{-1}$ for both withdrawal and infusion. **(c)** Drift measured from 149 tests across different capillaries and flow rates shows a small and consistent value of $2.3 \pm 0.3$ pC-s$^{-1}$.

**Supplementary Note 3: High Speed Imaging and Post Processing**

To extract quantitative information on water interface displacement from high-speed imaging data, we developed a custom MATLAB script that processes 8-bit grayscale TIFF images recorded at known frame rates.

The image processing algorithm is based on background subtraction and binary segmentation to extract the vertical motion of the water interface. A static background was generated by averaging the first five frames, assuming negligible initial movement. Each subsequent frame was then subtracted from this background to isolate motion-induced changes. Pixel contrast was enhanced, and a fixed threshold was applied to produce a binary mask of moving regions. Morphological operations (closing and hole filling) and area-based filtering were used to suppress noise and retain physically meaningful features. The vertical displacement of the interface in each frame was quantified as the maximum vertical extent of the foreground region. This simple yet robust method effectively suppresses noise and captures the spatiotemporal dynamics of the interface (see Figure S3).

Each displacement curve was then smoothed using a median filter to suppress local spikes while preserving overall trends. To improve physical interpretability, non-physical fluctuations prior to motion onset and after peak displacement were suppressed by isolating the active phase and enforcing a monotonic (non-decreasing) constraint. All traces were temporally aligned using the threshold-crossing point marking the onset of motion, and interpolated onto a common time grid for ensemble averaging (Figure S4).

High-speed imaging revealed that at release rates of 1 and 10 mL·min$^{-1}$, the 50 μL liquid volume was discharged from the capillary in two distinct droplets (Supplementary Movies 1 and 2). After detachment of the first droplet, a noticeable delay preceded the formation of the second droplet. This delay is likely attributable to the time required to re-establish sufficient pressure to overcome the Laplace pressure associated with high interfacial curvature. In the displacement-time curves, this delay manifests as a plateau during which the liquid interface remains stationary. Notably, the standard deviation (shaded region in Figure S4 and Figure S5a) increases markedly at the beginning of the second droplet's movement due to variations in the onset timing across trials, rather than sudden displacement events.

It is also noteworthy that although two droplets are still formed at 10 mL·min$^{-1}$, the interval between them is less than one second. Consequently, with the charge measured at 1 Hz, the two-step change in the charge signal cannot be resolved, resulting in a single observed step. In contrast, at 1 mL·min$^{-1}$, the two-step charge response is clearly detected (Supplementary Note 2 and Figure S2b).

The resulting displacement-time signals were further processed to derive velocity and acceleration. To extract smooth and physically meaningful motion profiles, a cubic smoothing spline was applied to the ensemble-averaged displacement data using a high smoothing parameter (p = 0.999). This produced a continuous trajectory that was differentiated numerically to obtain velocity and acceleration. We evaluated different smoothing levels (p = 0.90, 0.95, and 0.999) and compared them to Savitzky–Golay filters (5th-order polynomials with window sizes of 7, 15, and 37). While Savitzky–Golay filtering preserved the displacement signal (Figure S5a), its derivatives were noisy and oscillatory due to poor high-frequency noise suppression (Figure S5b-c)[1]. In contrast, the cubic spline provided smooth, $C^2$-continuous curves that yielded cleaner derivative profiles[2]. However, excessive smoothing (p ≤ 0.95) overly dampened sharp features, reducing peak magnitudes and distorting motion onset. A setting of p = 0.999 offered an optimal balance, preserving key kinematic features while reducing noise. Although this may slightly underestimate peak velocity and acceleration, uniform application across all trials ensures valid relative comparisons.

All experiments were conducted using a fixed liquid volume of 50 μL and capillaries of identical diameter, resulting in a consistent travel distance of approximately 40 mm. Under these controlled conditions, maximum velocity and maximum acceleration were used as key metrics to compare the dynamics of liquid motion. The corresponding velocity and acceleration profiles, derived from smoothed displacement data, are presented in Figure S6. The extracted peak values are summarized in Table S2.

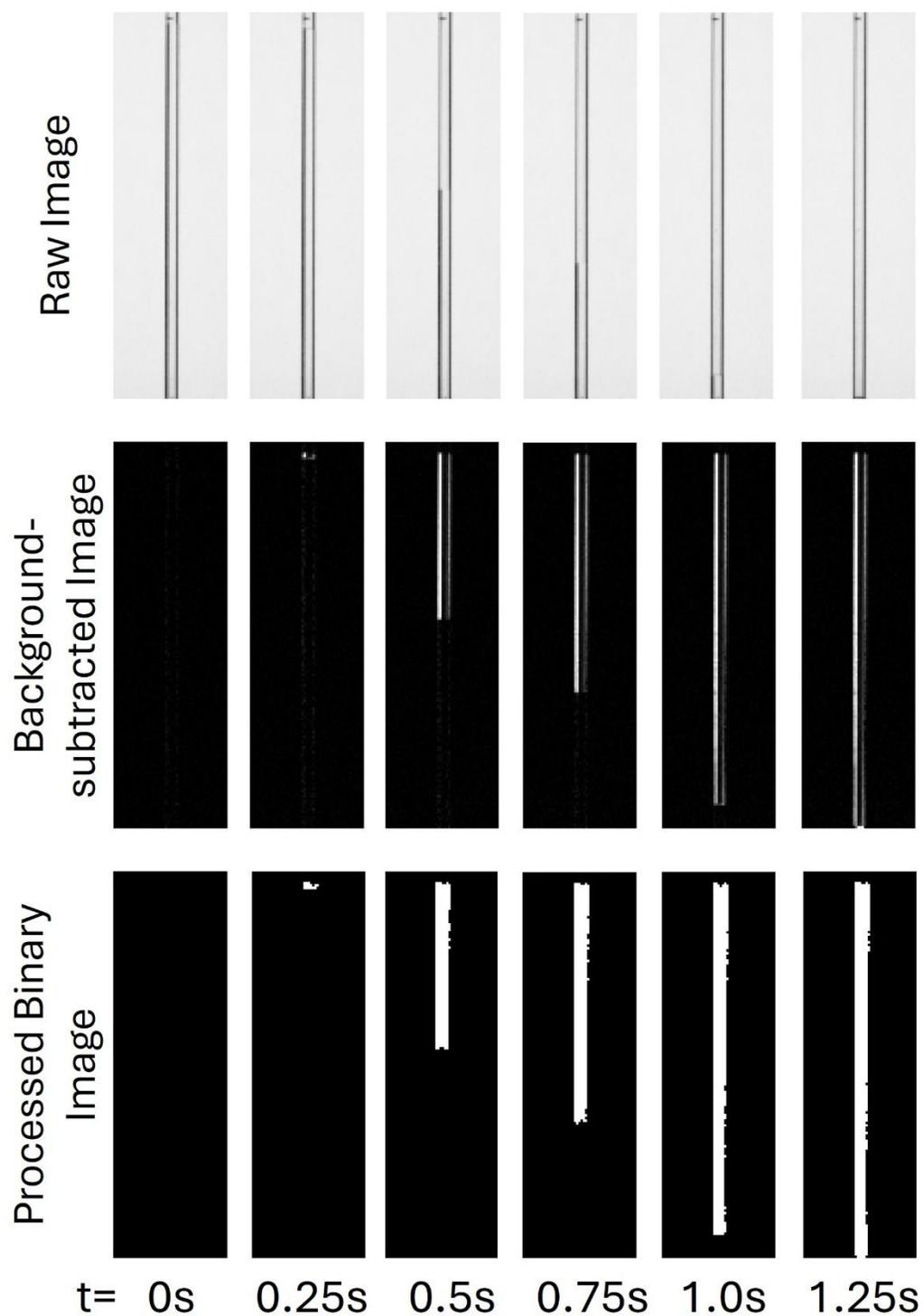

**Figure S3.** Image processing steps for tracking vertical water interface motion over time (t = 0–1.25 s). Top row: raw high-speed images. Middle row: background-subtracted images using an average of initial frames. Bottom row: processed binary images after contrast enhancement, thresholding, morphological filtering, and noise removal. The interface displacement is quantified by the maximum vertical extent of the detected regions. Example shown: ODTS-treated capillary, release rate = 10 mL-min$^{-1}$.

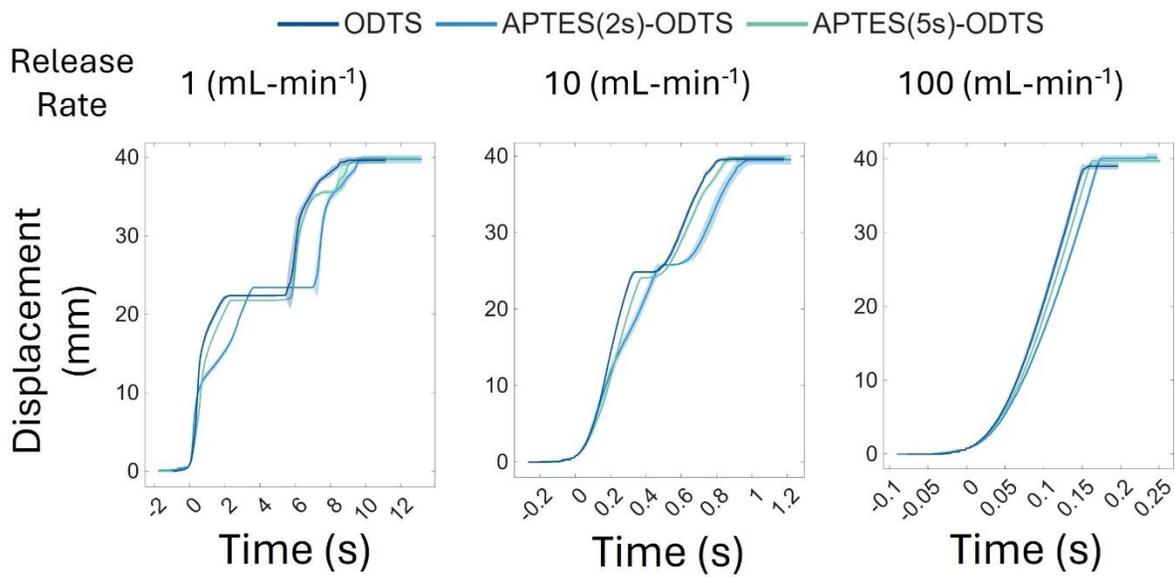

**Figure S4.** Displacement profiles for three surface treatments (ODTS, APTES(2s)-ODTS, and APTES(5s)-ODTS) under different release rates (1, 10, and 100). Shaded regions represent the standard deviation across repeated trials for displacement. The displacement curves exhibit minimal variability, highlighting the consistency of measurements and the robustness of the post-processing method.

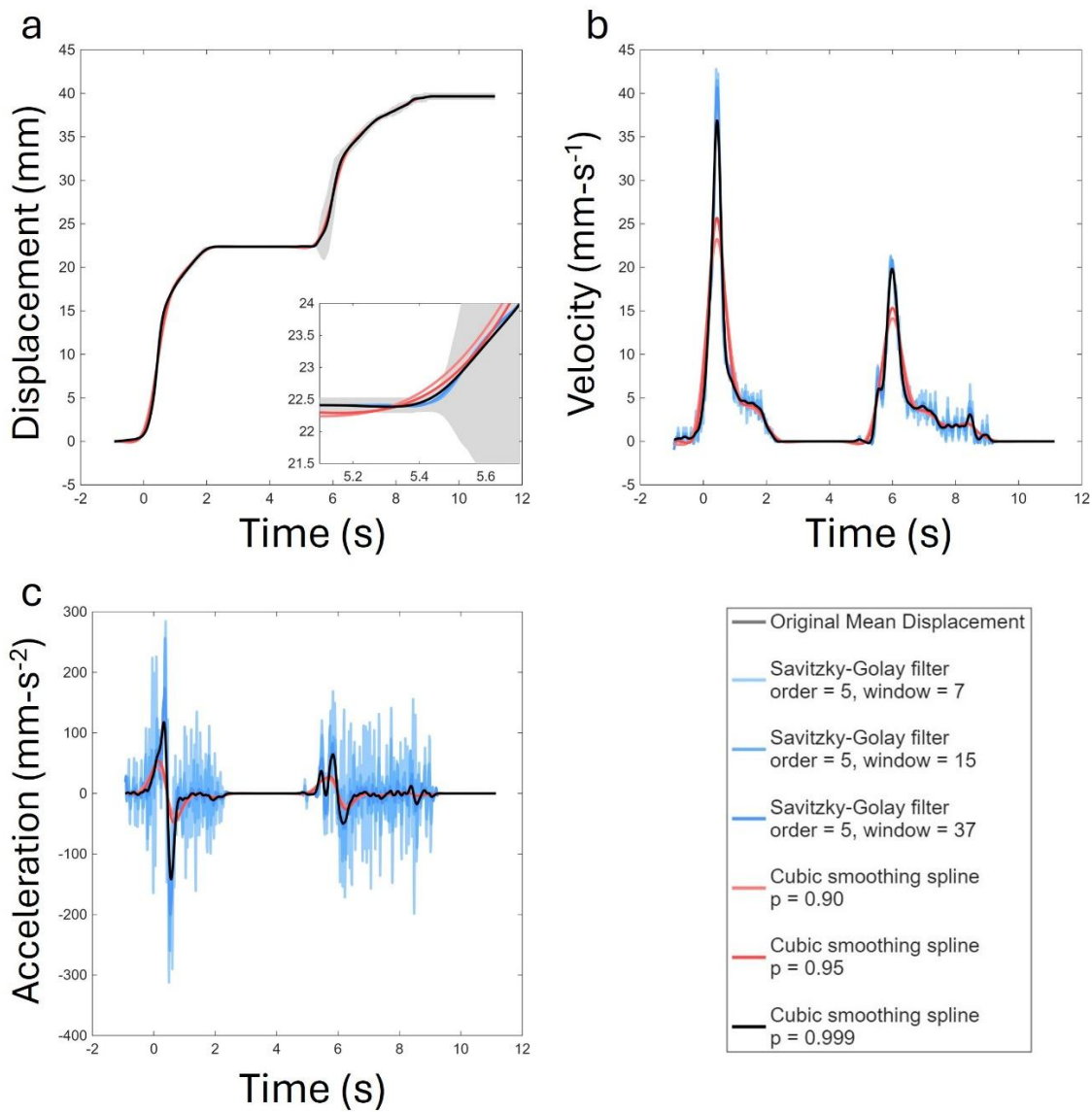

**Figure S5.** Comparison of smoothing methods applied to displacement (a), velocity (b), and acceleration (c) data. Displacement curves were smoothed using cubic splines with smoothing parameters p = 0.90, 0.95, and 0.999 (red and black lines), and Savitzky–Golay filters (5th-order polynomial, window sizes 7, 15, and 37; blue lines). The original unsmoothed data are shown in gray. In (a), Savitzky–Golay filters closely match the original displacement data, while lower spline parameters (e.g., p = 0.90) introduce artifacts such as unrealistic rollback near the onset of motion (see inset). In (b) and (c), Savitzky–Golay filters fail to suppress noise in velocity and acceleration, whereas cubic spline smoothing (especially at p = 0.999) provides cleaner, more physically plausible profiles. Example shown: ODTS-treated capillary, release rate = 1 mL-min$^{-1}$.

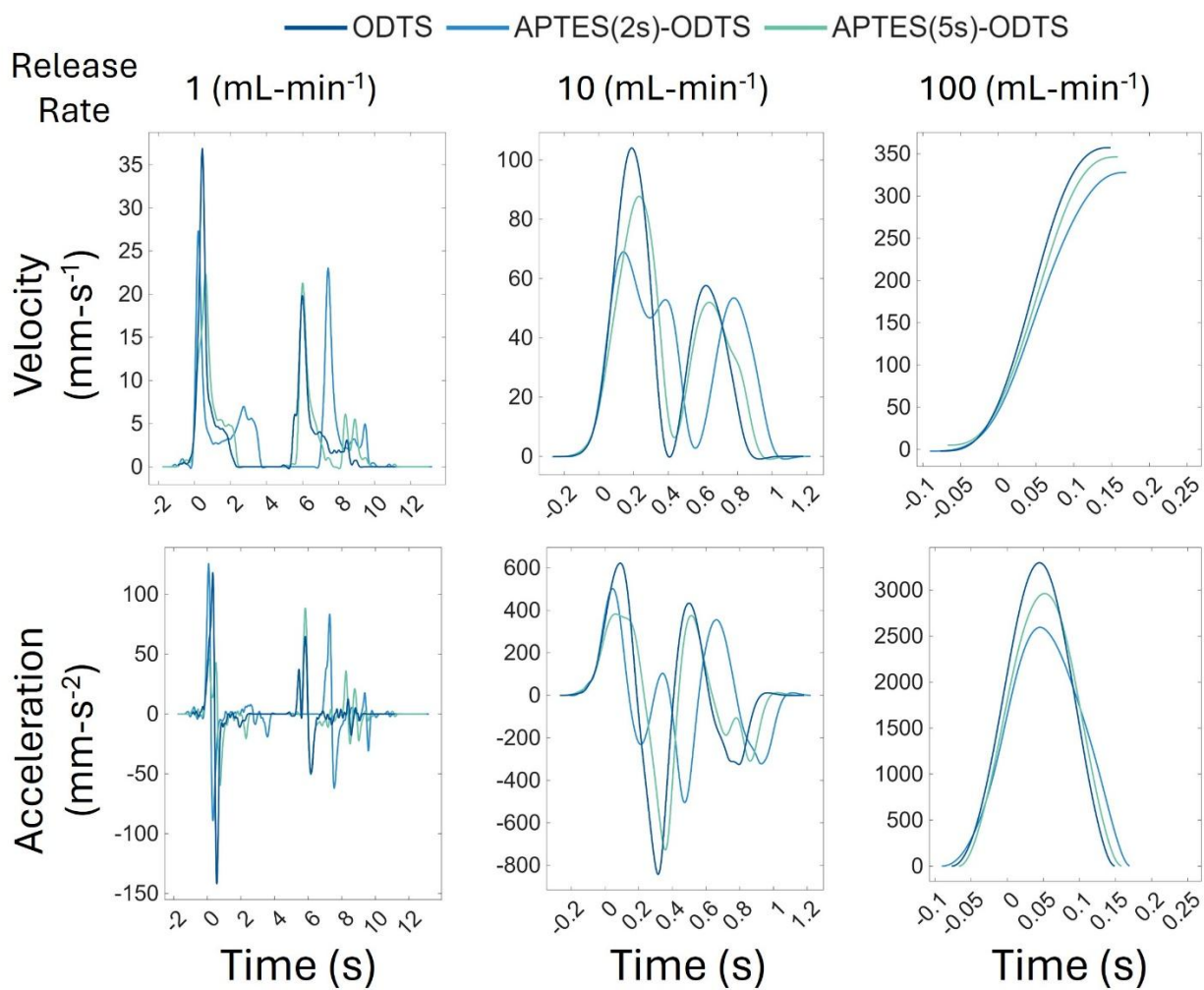

**Figure S6.** Velocity and acceleration profiles of liquid motion under different release rates (1, 10, and 100 mL-min$^{-1}$) for ODTS-, APTES(2s)-ODTS-, and APTES(5s)-ODTS-treated surfaces. The data were smoothed and differentiated to extract maximum values of velocity and acceleration, as summarized in Table S2.

**Table S2.** Maximum Velocity and Acceleration at Different Release Rates for Treatments

| Release Rate (mL-min$^{-1}$) | | ODTS | APTES(2s)-ODTS | APTES(5s)-ODTS |
|---|---|---|---|---|
| 1 | $V_{max}$ (mm-s$^{-1}$) | 36.9 | 27.3 | 22.4 |
|   | $A_{max}$ (mm-s$^{-2}$) | 118 | 126 | 88 |
| 10 | $V_{max}$ (mm-s$^{-1}$) | 104.0 | 69.0 | 87.7 |
|    | $A_{max}$ (mm-s$^{-2}$) | 622 | 502 | 383 |
| 100 | $V_{max}$ (mm-s$^{-1}$) | 357.3 | 328.0 | 346.3 |
|     | $A_{max}$ (mm-s$^{-2}$) | 3296 | 2594 | 2962 |

## Supplementary Note 4: Limitations of Classical Streaming Current Models in Finite-Volume Capillary Flows

A classical model for streaming current usually combined a set of charge density $\rho_e$ and flow velocity $v$ in a straight cylindrical pipe of radius $R$. It also includes the following assumptions:

- Charge density $\rho_e$, usually described by Poisson-Boltzmann Equation, depends only on the radial coordinate $r$, i.e., $\rho_e = \rho_e(r)$, and is time-invariant.
- Flow velocity $v$ may vary with both r and time t, i.e., $v = v(r, t)$.
- $\rho_e(r)$ and $v(r, t)$ are decoupled.

The current $I(t)$ through the pipe's cross-section at time $t$ is given by:

$$I(t) = \int_0^R \rho_e(r) \cdot v(r, t) \cdot 2\pi r \, dr$$

Therefore, the total charge $Q$ transported over time interval $[0, T]$ is:

$$Q = \int_0^T I(t) \, dt = \int_0^T \left[ \int_0^R \rho_e(r) \cdot v(r, t) \cdot 2\pi r \, dr \right] dt$$

By interchanging the order of integration, we get:

$$Q = \int_0^R \rho_e(r) \cdot \left[ \int_0^T v(r, t) \, dt \right] \cdot 2\pi r \, dr$$

Define the cumulative displacement length of fluid at radial position r as:

$$L(r) := \int_0^T v(r, t) \, dt$$

The total charge can be rewritten as:

$$Q = \int_0^R \rho_e(r) \cdot L(r) \cdot 2\pi r \, dr \quad \text{(S5a)}$$

In our case, the total volume $V$ of liquid dispensed is fixed, thus:

$$V = \int_0^T \left[ \int_0^R v(r, t) \cdot 2\pi r \, dr \right] dt = \int_0^R L(r) \cdot 2\pi r \, dr = constant \quad \text{(S5b)}$$

Combining equations (S5a) and (S5b), we found:

$$\frac{Q}{V} = \frac{\int_0^R \rho_e(r) \cdot L(r) \cdot 2\pi r \, dr}{\int_0^R L(r) \cdot 2\pi r \, dr} \quad \text{(S5c)}$$

As shown in Equation (S5c), the ratio of total charge to total volume is expressed on the right-hand side of the equation. In our experiments, the motion of a finite liquid volume within the capillary is constrained by the liquid–gas interfaces at both ends. From a macroscopic perspective, the displacement length $L(r)$ remains independent of the capillary radius $r$ and is equal to the liquid travel length, $L(r) = L_0$. This outcome is a direct consequence of the assumption that the charge density is decoupled from the flow velocity: the total amount of charge contained within the finite volume is constant and is entirely displaced along with the fluid. Therefore, such models are not directly applicable to describing the phenomena observed in our experiments. A more appropriate model for the present study would likely require a refined treatment of charge adsorption at the interface or a more detailed characterization of the flow field.